\documentclass[sigconf]{acmart}
\usepackage{colortbl}
\usepackage{amsmath}
\usepackage{xcolor}
\usepackage{graphicx}
\usepackage{subcaption}
\usepackage{mwe}
\usepackage{adjustbox}
\usepackage{multirow}
\colorlet{tablerowcolor}{gray!10} 
\newcommand{\rowcol}{\rowcolor{tablerowcolor}} %

\AtBeginDocument{%
 }

\copyrightyear{2024}
\acmYear{2024}
\setcopyright{rightsretained}
\acmConference[CIKM '24]{Proceedings of the 33rd ACM International Conference on Information and Knowledge Management}{October 21--25, 2024}{Boise, ID, USA}
\acmBooktitle{Proceedings of the 33rd ACM International Conference on Information and Knowledge Management (CIKM '24), October 21--25, 2024, Boise, ID, USA}
\acmDOI{10.1145/3627673.3679172}
\acmISBN{979-8-4007-0436-9/24/10}

\makeatletter
\gdef\@copyrightpermission{
  \begin{minipage}{0.3\columnwidth}
   \href{https://creativecommons.org/licenses/by/4.0/}{\includegraphics[width=0.90\textwidth]{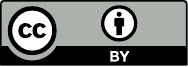}}
  \end{minipage}\hfill
  \begin{minipage}{0.7\columnwidth}
   \href{https://creativecommons.org/licenses/by/4.0/}{This work is licensed under a Creative Commons Attribution International 4.0 License.}
  \end{minipage}
  \vspace{5pt}
}
\makeatother

\begin{document}
\title{An Evaluation Framework for Attributed Information Retrieval using Large Language Models}

\author{Hanane Djeddal}
\email{hanane.djeddal@irit.fr}
\affiliation{%
\institution{Université Paul Sabatier, IRIT}
  \postcode{31000}
  \city{Toulouse}
  \country{France}}
  \affiliation{\institution{Ecovadis}
  \postcode{75116}
  \city{Paris}
  \country{France}}

\author{Pierre Erbacher}
\author{Raouf Toukal}
\author{Laure Soulier}
\email{pierre.erbacher@isir.upmc.fr}
\email{raouf.toukal@isir.upmc.fr}
\email{laure.soulier@isir.upmc.fr}
\affiliation{%
  \institution{Sorbonne Université, CNRS, ISIR}
  \postcode{F-75005 }
  \city{Paris}
  \country{France}
}
\author{Karen Pinel-Sauvagnat}
\email{Karen.Sauvagnat@irit.fr}
\affiliation{%
  \institution{Université Paul Sabatier, IRIT}
  \postcode{31000}
  \city{Toulouse}
  \country{France}
}
\author{Sophia Katrenko}
\email{skatrenko@ecovadis.com}
\affiliation{%
  \institution{Ecovadis}
  \postcode{75116}
  \city{Paris}
  \country{France}}
  
\author{Lynda Tamine}
\email{lechani@irit.fr}
\affiliation{%
  \institution{Université Paul Sabatier, IRIT}
  \postcode{31000}
  \city{Toulouse}
  \country{France}
}

\renewcommand{\shortauthors}{Hanane Djeddal et al.}

\begin{abstract}
  

  With the growing success of Large Language models (LLMs) in information-seeking scenarios, search engines are now adopting generative approaches to provide  answers along with in-line citations as attribution. While existing work focuses mainly on attributed question answering, in this paper, we target information-seeking scenarios which are often more challenging due to the open-ended nature of the queries and the size of the label space in terms of the diversity of candidate-attributed answers per query. We propose a reproducible framework to evaluate and benchmark attributed information seeking, using any backbone LLM, and different architectural designs: (1) Generate (2) Retrieve then Generate, and (3) Generate then Retrieve. Experiments using HAGRID, an attributed information-seeking dataset, show the impact of different scenarios on both the correctness and attributability of answers.

\end{abstract}

\ccsdesc[500]{Information systems~Information retrieval}

\keywords{Large Language Models, Answer generation, Attributed information seeking}

\maketitle
\section{Introduction}
\label{intro}
Large Language Models (LLMs) \cite{NEURIPS2020_1457c0d6, achiam_gpt-4_2023}  have shown great potential in information-seeking scenarios. This has led to a paradigm shift from document ranking to natural language answer generation to address the user's query. While this new approach helps overcome several challenges previously faced by traditional search solutions, it also brings new ones to light. One such challenge stems from LLMs' tendency to generate unfaithful text often referred to as “hallucination” \cite{peskoff-stewart-2023-credible}. This problem becomes critical in information-seeking and knowledge-intensive scenarios where users expect contextually rich, up-to-date, and reliable responses to their queries \cite{li_survey_2023}. Attribution has emerged as a new solution where texts are generated by LLMs with supporting sources called \textit{attributions},  allowing users to fact-check the model's claims \cite{Bohnet_2022, gao-etal-2023-enabling}. 
Current works on attributed search often leverage existing Question Answering (QA) datasets to benchmark and evaluate the performance of different attribution methods \cite{gao-etal-2023-enabling, Bohnet_2022, gao-etal-2023-rarr}. However, attributed information seeking is more challenging due to the open-ended nature of the query and the scale of the label space of the query and its answer, as well as the answer and its attribution \cite{Bohnet_2022}. A query can have multiple possible answers, supported by multiple possible sources. These aspects are important to consider for the evaluation of attributable information-seeking approaches, but QA frameworks do not inherently account for them \cite{Bohnet_2022,gao-etal-2023-enabling}. 
\begin{figure*}[t]
    \centering
      \includegraphics[width=\linewidth]{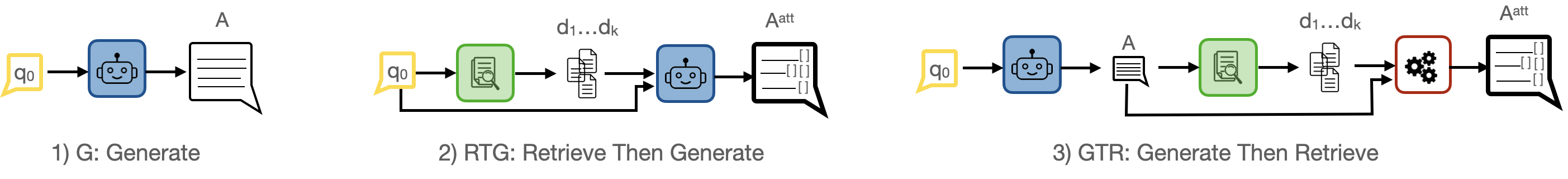}
      \vspace{-0.8cm} 
    \caption{LLM-based scenarios for IR with attribution.}
    \label{fig:approches}
\end{figure*}
While there has been a continuous rise in the number of contributions in this field \cite{li_survey_2023, hu2024benchmarking, asai2023selfrag}, there is still a critical lack of open-source frameworks useful for designing, developing, and evaluating attributable information-seeking systems. This paper tackles this gap with the hope of fostering research in this area. While our proposed framework can be used for any dataset targeting an information-seeking task, we perform experiments and report results obtained using the recently designed HAGRID  dataset \cite{kamalloo_hagrid_2023}, establishing its baselines following three architectures commonly used for attribution: 1) Generate (\textbf{G}) \cite{sun2023recitationaugmented,weller-etal-2024-according, NEURIPS2020_1457c0d6} where the model is prompted to generate an answer with or without its attribution; 2) Retrieve then Generate (\textbf{RTG}) \cite{pmlr-v119-guu20a,pmlr-v162-borgeaud22a,asai2023selfrag} where the system first retrieves a list of documents relevant to the user query and then generates an answer based on the query and the relevant documents, and 3) Generate then retrieve (\textbf{GTR}) \cite{gao-etal-2023-rarr, 10.1145/3624918.3625336} where attribution is added in a post-generation stage.  While our framework is also agnostic to backbone LLMs used, we report experiment results using Zephyr$_{3b}$\footnote{https://huggingface.co/stabilityai/stablelm-zephyr-3b}, a 3 billion parameter finetuned version of mistral \cite{jiang2023mistral}. Regarding evaluation, we measure both the correctness of the answer and the quality of the attribution using a set of different metrics reported in the literature \cite{Bohnet_2022, gao-etal-2023-enabling, gao-etal-2023-rarr}, relying on either human-built golds or automatically-built references. In the latter case, we use evaluation metrics based on Natural Language Inference (NLI) models \cite{Xie2021FactualCE} which has been found to correlate well with human judgments \cite{Bohnet_2022} making attribution evaluation and benchmarking possible without the need for gold citation annotation.

The contributions of this paper are as follows: 1)  we provide an open-source framework that implements key architectures using any backbone LLM, and provides several automatic evaluation metrics for attributed information seeking using different references either automatically built-up or human-assessed;  2) we perform a systematic analysis of the different proposed architectures and key parameters and components, including the impact of retrieval on generation;  3)  as of the time of this writing and to our knowledge, this is the first benchmarking of the dataset HAGRID, that could be easily used with new datasets adequate for attributable generative information seeking scenarios. All implementations of scenarios and metrics as well as the generated outputs are available online\footnote{\url{https://github.com/hanane-djeddal/Attributed-IR}}.

\section{Experimental design}
\label{exp_design}
\subsection{Task formulation}


The task of attributed information seeking is formalized as follows: given a user query $q_0$ and a corpus of document passages $D$, the goal is to generate an attributed answer  $A^{att}$ in response to the query $q_0$.  $A^{att}$ is a set of $n$ distinct statements (i.e. sentences) $a_i$,  where each statement $a_i$ has in line citations referring to a set $C_i =\{c_{i1},c_{i2,...}\}$ of documents that support the statement $a_{i}$,  with $c_{ij}\in D$. 
Following these notations, an attributed answer is represented as $A^{att} = \{(a_1,C_1), (a_2,C_2),...(a_n,C_n)\}$, while an answer without attribution is represented as  $A = \{a_1, a_2, ..., a_n\}$. 



\begin{table}
  \caption{Statistics on HAGRID dataset. \#Q is the number of queries. \#A the number of answers. \#Inform is the number of informative answers. \#Attrib is the number of  Attributable answers. avg \# P and avg \# Cit are resp. the average number of relevant passages per query and citations by answer. }
  \vspace{-0.2cm}
  \label{tab:stats_hagrid}
  \begin{tabular}{ccccccc}
    \toprule
    Split  & \#Q & \#A & \#Inform &	\#Attrib & avg \#P & avg \#Cit\\
    \midrule
    Train & 1922&	3214	& 3214	&754 &  2.73& 2\\
     Dev & 716&	1318	&1157&	826 &2.91   & 2.5\\
  \bottomrule
\end{tabular}
\vspace{-0.2cm}
\end{table}
\subsection{Dataset}

We use the HAGRID dataset \cite{kamalloo_hagrid_2023}, built on top of the MIRACL English subset \cite{zhang2022making} where 
each query and its human-assessed relevant documents are fed to an LLM (GPT-3.5 turbo) to generate answers. Human annotators evaluate the LLM-generated answers according to two criteria: 1) \textit{informativeness} refers to the correctness of the answer statements regarding the query; 2) \textit{attributability} refers to the correspondence between the answer statements and the gold citations. Statistics on the dataset are shown in Table \ref{tab:stats_hagrid}.
Unlike other datasets used for attributed generative question-answering \cite{Bohnet_2022,gao-etal-2023-enabling},  the HAGRID dataset has three main peculiarities that motivate us to use it for designing our benchmark: 1) its core dataset MIRACL targets an-ad-hoc information retrieval task; 2) it provides a gold set of answers with attributions and human annotations w.r.t to both informativeness of answer statements $a_i$ and their attribution quality; and  3) despite its great usefulness for the IR and NLP communities to evaluate attributable generative information-seeking models, it is not so far used for a baseline evaluation with  LLM-based approaches.  

\subsection{Scenarios} \label{approaches}
Here, we describe the three main architectures used for answer generation with attribution in literature \cite{Bohnet_2022, gao-etal-2023-enabling, liu-etal-2023} depicted in Figure \ref{fig:approches}.
We report our experiment results using Zephyr$_{3b}$ but results with other LLMs are provided with the code demo \footnote{\url{https://github.com/hanane-djeddal/Attributed-IR}}. 

\noindent (1) \textbf{Generate (G)}: this scenario  directly generates an answer $A$ using an LLM in response to an input query $q_0$ in a closed-book fashion. Since the model solely depends on its pre-training data, we stick to the simple case of generating an answer $A$ without citations. 
    
\noindent (2) \textbf{Retrieve Then Generate (RTG)}: this scenario first retrieves a list $P=\{d_1,..,d_k\}$ of $k$ candidate supporting documents to the query $q_0$ from the corpus $D$, which are then fed to the LLM with $q_0$ to generate an attributed answer $A^{att}$.  
We investigate two scenarios: 
    
    $\bullet$ \textit{Vanilla retrieval (vanilla)}: we use a two-stage ranker based on BM25 \cite{bm25} and MonoT5 \cite{nogueira-etal-2020-document} to retrieve the list $P$ of relevant documents to  query $q_0$ used to generate the attributed answer $A^{att}$. 
    
$\bullet$ \textit{Retrieval with query generation (query-gen)}: we propose a new variant of the RTG architecture where we leverage the LLM's parametric memory 
to guide the retrieval towards documents that align with the model's internal knowledge using query reformulations. We hope to bridge the gap between the model's stored knowledge and externally retrieved content \cite{li_survey_2023, Wu2024HowFA, Zhang2023TheKA}.  To do so, given the user query $q_0$, the LLM first generates a list of subqueries $Q=\{q_1,..,q_m\}$  based on $q_0$. For each generated query $q_t \in Q$ as well as the user query $q_0$, we retrieve $k$ candidate relevant documents $P_t = \{d_{t1}, d_{t2},..,d_{tk}\}$  to the query $q_t$, for $t=0,..,m$, using the retrieval model used in the \textit{vanilla} scenario. Then, we aggregate using a data fusion technique (Cf. Section 2.4) the different lists of documents $P_t$ with $t=0,..,m$ to produce one final list of candidate documents $P=\{d_1,..,d_k\}$ which is then fed to the LLM to generate a grounded attributed answer $A^{att}$. 

\noindent (3) \textbf{Generate Then Retrieve (GTR)}: This scenario first generates an answer $A$ without citations, which is then used to identify relevant documents that support it. More specifically, to generate inline citations along answer $A$, we split it into statements $a_i$ for which the retriever returns a ranked list $P_i$ of $k$ relevant documents $P_i = \{d_{i1}, d_{i2},..,d_{ik}\}$ that we consider as the associated citations to 
build the attributed answer $A^{att}$. 
Similar to \textit{vanilla} and \textit{query-gen} scenarios, we use BM25+MonoT5 for the retrieval.







\subsection{Variants and baselines}
To better understand the impact of retrieval on performances in the RTG scenario, we establish a baseline\textbf{ RTG-gold} where the LLM generates the answer using the gold documents. To evaluate the quality and the impact of our new scenario of generating $m$ queries (\textbf{RTG-query-gen}), we test different ranking aggregation methods from the literature: 1) \textbf{CombSum} \cite{Fox1993CombinationOM}, 2) \textbf{CombSumMNZ} \cite{Fox1993CombinationOM} and 3) \textbf{PM2} \cite{10.1145/2348283.2348296}. 
We also define additional baseline methods: 4) \textbf{Sort}: sorting the documents from the various lists by their normalized MonoT5 score, and 5) \textbf{Rerank}: reranking the documents from the different lists using MonoT5. 
We also report the best possible performance for query-gen (\textbf{MAX}) by evaluating the retrieval results of each generated query separately and choosing the best score for each metric.  For the answer generation, we use the documents issued from \textbf{Rerank}.

\subsection{Evaluation metrics} \label{eval_metrics}

We consider in our framework two main categories w.r.t the output structure including both answer statements and citations. 

\vspace{-0.2cm}
\paragraph{\textbf{Answer correctness.}}
Contrary to QA settings where answers can be short and exact, in information-seeking scenarios,  answers tend to be long and can vary in form. Therefore we use different n-gram-based and semantic-based metrics: ROUGE \cite{lin2004rouge}, BLEU \cite{papineni2002bleu}, and BertScore \cite{zhang2019bertscore} to evaluate the accuracy of the answer based on the gold answer. Since a query can have multiple possible answers in the ground truth, we report for each metric, the best score calculated by comparing the generated answer with each candidate answers.

\vspace{-0.2cm}
 
\paragraph{\textbf{Citation quality.}} We apply different metrics to measure either the answer statements' attribution w.r.t. the associated citations in the answer or the gold citations. 

$\bullet$  \textbf{AutoAIS} \cite{gao-etal-2023-rarr}: This metric was originally defined for the task of attributed QA where the citations are not in line with the answer but reported in a separate list associated with the answer as a whole. In our case, each statement $a_i$ has its own set of citations $C_i$ in line with the answer $A^{att}$. Thus, we adapt this metric: 
        \begin{equation}
        AutoAIS_{cit}(A^{att})= avg_{(a_i,C_i)\in A^{att}} [argmax_{c_{ij} \in C_i} (NLI(c_{ij},a_i))]
        \end{equation}
        where \textit{NLI (premise, hypothesis)} is the output of an NLI classifier (1 if the premise entails the hypothesis, and 0 otherwise). This first application of AutoAIS determines whether a statement is attributed to at least one of its associated citations.\\
    Moreover, to investigate to which extent an answer $A^{att}$ is attributable to the retrieved documents  $P= \{d_1,..,d_k\}$ fed to the LLM in the \textbf{RTG} scenario, we introduce a variant of the AutoAIS metric, $AutoAIS_{pssg}(A^{att})$,  as follows: 
        \begin{equation}
        AutoAIS_{pssg}(A^{att})= avg_{a_i\in A} [argmax_{d_l \in P} (NLI(d_l,a_i))]
        \end{equation}
$\bullet$  \textbf{Citation Recall and Precision}: Facing the lack of gold citations, the framework ALCE \cite{gao-etal-2023-enabling} introduced a citation recall and citation precision using the automatic binary score (0,1) of an  NLI classifier between the statements and the citations. Citation recall determines if a statement $a_i$ is entirely supported by its cited documents by examining the NLI score between the statement and the concatenation of its cited documents denoted $concat(C_i)$ i.e. by computing $NLI(concat(C_i), a_i)$. Citation precision measures to which extent a citation $c_{ij}$ is irrelevant by checking whether $NLI(c_{ij}, a_i)=0$ and if it can be removed without affecting the attribution score of its statement $a_i$ such that $NLI(concat(C_i\setminus c_{ij}), a_i)=1$. We use the implementation provided in the ALCE framework\footnote{https://github.com/princeton-nlp/ALCE/tree/main} and refer to them using NLI\_prec and NLI\_rec in the rest of the paper. \\

\vspace{-0.3cm}
$\bullet$ \textbf{Citation Overlap}: Since HAGRID provides gold answers $A^{att}$ with gold citations $\mathcal{C}_{gold}$, we perform a more direct evaluation of citation attributability by computing for each query the overlap between the gold citations  $\mathcal{C}_{gold}$ and the whole citations in the generated answer $\mathcal{C}_{gen}$. We define citation overlap precision and recall for answer $A^{att}$ to query $q$ as:
\begin{equation}
        Overlap (A^{att}) = \frac{|\mathcal{C}_{gold} \cap \mathcal{C}_{gen}| }{|X|}
        \end{equation}
    Where $\mathcal{C}_{gen}=\bigcup_{(a_i,c_i)\in A^{att}} C_i$, $|.|$ is the cardinality of the set.  $X$ equals to $\mathcal{C}_{gen}$ or   $\mathcal{C}_{gold}$ for computing the overlap precision or overlap recall, resp..

\section{Experiments}
\label{exp_res}

\begin{table*}
\small
  \caption{Main results for G, RTG, and GTR scenarios using the HAGRID dataset. \textit{Gold answer} reports the citation quality in the gold answer using automatic metrics. Since the scenario \textit{G} does not include citations, only answer correctness is evaluated. The generated simple answer in scenario \textit{GTR} and \textit{G} are the same, thus the answer correctness measures are equal. The best performance measures are in bold, and the second best is underlined.}
  \vspace{-0.3cm}
  \label{tab:main_results}
  \begin{tabular}{c|ccccccc|cccccl}
    \toprule
     \multirow{3}{4em}{} & \multicolumn{7}{c|}{Correctness} &  \multicolumn{6}{c}{Citations}\\
    \cline{2-14}
     &\multicolumn{1}{c}{BLEU} & \multicolumn{3}{c}{ROUGE-L} &  \multicolumn{3}{c|}{BertScore} & \multicolumn{2}{c}{Overlap} &\multicolumn{2}{c}{AutoAIS} & \multicolumn{2}{c}{NLI prec/rec}  \\
    \cline{2-14}
     &  &  Prec. & Rec. & F-score  & Prec. & Rec. & F-score& Prec. & Rec.& Cit. & Pssg. & Prec. & Rec. \\
    \midrule
   Gold answer  & & - & -& - &-&- &-&-&-&87.97& 89.21&83.65&79.80\\
    G  & 11.06 & 30.41  & 46.08 & 31.58 & 87.88 & 90.02 & 88.87 &-&-&-  &-&-&-\\
    RTG - gold & \textbf{28.22} & \textbf{44.00}  & \textbf{63.81} & \textbf{46.72} & \textbf{90.02} &\textbf{ 93.36} &\textbf{ 91.69} &\textbf{75.29}&\textbf{68.89}& \textbf{42.81} & \textbf{80.67} &56.55&\underline{42.31} \\
    RTG - vanilla  & \underline{18.44} & \underline{33.83} & \underline{56.40}& \underline{36.65} & \underline{87.94} & 91.52 & \underline{89.63} & \underline{36.17} & \underline{32.69} & 41.86 & 78.95 & \underline{57.90} & 41.63 \\ 
    RTG - query-gen  & 18.33 & 33.58 & 56.13 & 36.43 & 87.89 &\underline{91.55} & 89.62 & 35.89 & 32.46 & \underline{42.68} & \underline{80.10} & \textbf{59.59} & \textbf{42.48}\\    
    GTR  &11.06 & 30.41  & 46.08 & 31.58 & 87.88 & 90.02 & 88.87& 45.53 & 30.53 & 26.69 &26.69& 26.65&26.66\\   
  \bottomrule
\end{tabular}
\end{table*}

\subsection{Effectiveness results}

We report the main results of the different scenarios (cf. Section 2.3) in Table \ref{tab:main_results} using the evaluation metrics presented in Section \ref{eval_metrics}. For \textbf{RTG-vanilla} and \textbf{RTG-query-gen}, we fix the number of supporting documents to 2 for a fair comparison with the \textbf{RTG-gold} which relies on 2.7 gold documents on average (cf. Table 1). For the \textbf{GTR} scenario, we retrieve 1 document per statement to be close to the attributed answers in the ground truth including 2.5 citations per answer on average.
From a general point of view, we can see that \textbf{RTG} scenarios have the best overall performances which confirms the benefit of retrieval-augmented answer generation \cite{10.1145/3539618.3591687}. This can also be seen through the lower results obtained by the \textbf{GTR} scenario suggesting that using attribution at a post-generation stage can propagate the hallucinations in the generated answer making it hard to identify supporting documents. 
Moreover, \textbf{RTG-gold} sets a strong upper bound, showcasing the importance of the quality of the retrieved documents.  Among the evaluated \textbf{RTG} variants, \textbf{RTG-query-gen} has the best performance in citation quality regarding AutoAIS and NLI-based metrics. It also obtains higher results than the \textbf{RTG-gold} scenario regarding NLI-based metrics. This confirms our intuition that query reformulation helps in exhibiting parametric knowledge in LLMs towards information-seeking needs. 

\begin{figure}[t]
        \caption[  ]
        {\small Citations and Correctness metrics for varying values of the number of supporting documents in RTG-user-query scenario} 
        
        \begin{subfigure}[b]{0.2\textwidth}
            \includegraphics[width=\textwidth]{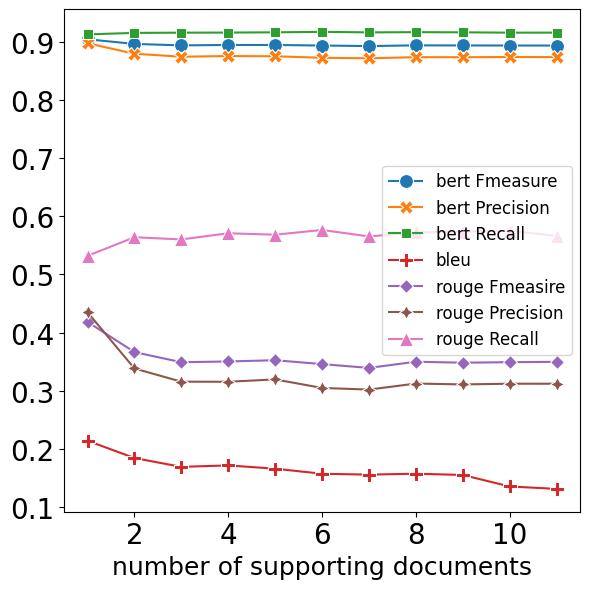}
            \caption[ Correctness metrics]%
            {{\scriptsize Correctness metrics}}    
            \label{fig:mean and std of net14}
        \end{subfigure}
        \hfill
        \begin{subfigure}[b]{0.2\textwidth}   
            \includegraphics[width=\textwidth]{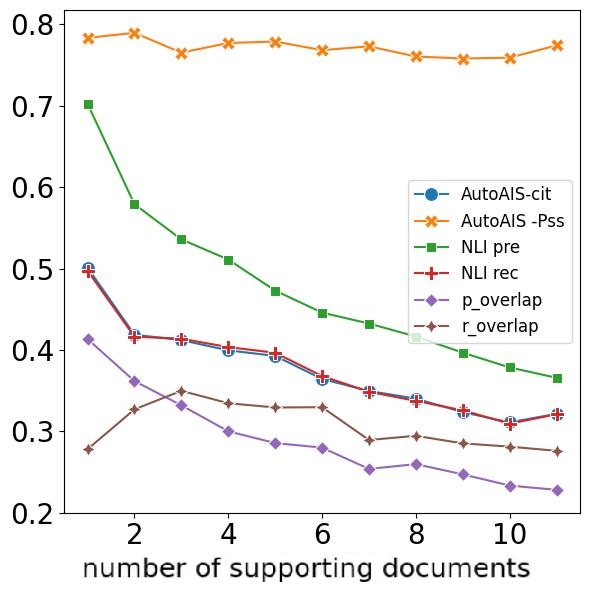}
            \caption[Citation metrics]%
            {{\scriptsize Citation metrics}}    
            \label{fig:mean and std of net34}
        \end{subfigure}
        \label{fig:varying_k}
        
\end{figure}

\begin{table}
\footnotesize
    \caption{\small Retrieval results for the RTG scenario. Best performances are in bold and the second best is underlined.}
    \vspace{-0.3cm}
\begin{adjustbox}{width=.5\textwidth,center}
    \begin{tabular}{p{0.6in}p{0.7in}@{}ccccc}
        \toprule
        \multirow{2}{4em}{Setting} & \multirow{2}{4em}{Agg} & \multicolumn{2}{c}{@1} & \multicolumn{3}{c}{@10}  \\
        \cline{3-7}
        && Precision & Recall  &Precision & Recall &nDCG \\
        \midrule
        user query & -&  \underline{48.04}& \textbf{23.29} &  \textbf{16.45} & 63.31& \textbf{53.21}\\
        \midrule
        \multirow{3}{4em}{query-gen} & sort& 36.17 & 17.87  & 13.84  & 57.03& 43.88\\

        & combSum& 47.07 &  22.37& 16.17  & 62.45& 51.39\\

         & combSum-MNZ&  45.39 &  21.22 &16.28  &61.98& 50.49\\

         
          & Rerank&46.79 & 22.83 & \underline{16.33} & \textbf{63.99}  &\textbf{53.21}\\
         
           & PM2 &\textbf{48.60} & \underline{23.16} & \underline{16.33} & \underline{63.43}  & \underline{52.83}\\
        \cline{2-7}
        &MAX &\textbf{ 66.34}  &  \textbf{32.42}  &  \textbf{18.66} & \textbf{72.69}  & \textbf{63.58}\\
        \bottomrule 
    \end{tabular}
 \label{tab:retrieval_results}
\end{adjustbox}
\vspace{-0.5cm}
\end{table}
We also note the difference between citation quality and faithfulness to input documents. AutoAIS$_{cit}$ scores can be lower than half the scores of AutoAIS$_{Pssg}$ across different scenarios. This means that while input documents can correctly support the sentences in the answer, they are not always correctly cited. 

\subsection{Complementary analysis}
\subsubsection{Impact of the number of supporting documents}
Figure \ref{fig:varying_k} shows the performances of the \textbf{RTG-vanilla} scenario when varying the number of supporting documents. Please note that trends are similar for the \textbf{RTG-query-gen} scenario. Regarding the correctness metrics (Figure 2(a)), the number of documents has a slight impact, with a small tendency towards the decrease of ROUGE metrics. Focusing on the citation metrics (Figure 2(b)), the number of documents seems to have little impact on AutoAIS$_{pssg}$, contrary to the other metrics which overall decline when increasing the number of documents. This suggests that the model may struggle to cite correctly when increasing the number of input documents, which may stem from a tendency to over-cite (increase in NLI precision without necessarily improving recall). We also note that the recall of citations-overlap peaks at around 2 documents which corresponds to the average number of gold documents in the dataset. 

\subsubsection{Impact of Retrieval}
 Using the generated queries in the \textbf{RTG-query-gen} scenario can improve the retrieval scenario, especially in citation quality. We, therefore, analyze in Table \ref{tab:retrieval_results} the impact of the different aggregation approaches over the generated queries on the retrieval effectiveness. 
 We can see that the re-ranking aggregation method (\textbf{Rerank}) seems to give the best retrieval score @10 though not improving much from the vanilla scenario. The best obtainable results using \textbf{query-gen} scenario can go up to 72\% recall @10, leaving room to investigate other aggregation methods. 
\subsubsection{Correlation between NLI and Annotations}
We evaluate the gold answers provided in HAGRID using the automatic citation metrics and examine how they correlate with the provided human annotation of attributability. We find that the Pearson correlation coefficient between $AutoAIS_{cit}$ and the attribution annotation is pretty modest at 0.36, similarly with citation precision and recall at $0.35$ and $0.36$ respectively. This is an interesting finding as previous work \cite{Bohnet_2022} found a strong correlation between human annotations and AutoAIS when using QA dataset for attribution, which testify to the complexity of attributed information-seeking and its distinction from attributed QA.

\section{Conclusion}
In this work, we propose an extensible open-source framework to benchmark attributable information-seeking using representative LLM-based approaches. 
We consider a set of up-to-date SOTA evaluation metrics to measure both the correctness of the generated answer and the quality of the in-line citations. Our work provides valuable baseline models and insights from empirical results and trends allowing to strengthen the maturity of research that targets faithful generative information-seeking.\\   

 \section{Acknowledgments}
 This work partially benefited from support from the French National Research Agency (Project GUIDANCE, ANR-23-IAS1-0003)



\bibliographystyle{ACM-Reference-Format}
\bibliography{main}










\end{document}